\documentclass[12pt,letterpaper]{article}

\usepackage{amsmath}
\usepackage{amsfonts}
\usepackage{amssymb}
\usepackage{graphicx}
\begin{document}


\begin{flushright} {\footnotesize MIT-CTP-3499}  \end{flushright}
\vspace{5mm} \vspace{0.5cm}
\begin{center}

\def\thefootnote{\fnsymbol{footnote}}

{\Large \bf Tilted Ghost Inflation} \\[1cm]
{\large Leonardo Senatore\footnote{This work is supported in part
by funds provided by the U.S. Department of Energy (D.O.E) under
cooperative research agreement DF-FC02-94ER40818}}
\\[0.5cm]

{\small
\textit{Center for Theoretical Physics, \\
Massachusetts Institute of Technology, Cambridge, MA 02139, USA}}
\end{center}
\vspace{.2cm}

 \vspace{.8cm}

\hrule \vspace{0.3cm}
{\small  \noindent \textbf{Abstract} \\[0.3cm]
\noindent In a ghost inflationary scenario, we study the
observational consequences of a tilt in the potential of the ghost
condensate. We show how the presence of a tilt tends to make
contact between the natural predictions of ghost inflation and the
ones of slow roll inflation. In the case of positive tilt, we are
able to build an inflationary model in which the Hubble constant
$H$ is growing with time. We compute the amplitude and the tilt of
the 2-point function, as well as the 3-point function, for both
the cases of positive and negative tilt. We find that a good
fraction of the parameter space of the model is within
experimental reach.

\vspace{0.5cm}  \hrule

\def\thefootnote{\arabic{footnote}}
\setcounter{footnote}{0}


\section{Introduction}

Inflation is a very attractive paradigm for the early stage of the
universe, being able to solve the flatness, horizon, monopoles
problems, and providing a mechanism to generate the metric
perturbations that we see today in the CMB \cite{Guth}.

Recently, ghost inflation has been proposed as a new way for
producing an epoch of inflation, through a mechanism different
from that of slow roll inflation \cite{Ark1,Ark2}. It can be
thought of as arising from a derivatively coupled ghost scalar
field $\phi$ which condenses in a background where it has a non
zero velocity:
\begin{equation}
<\dot{\phi}>=M^2 \rightarrow <\phi>=M^2t
\end{equation}
where we take $M^2$ to be positive.

Unlike other scalar fields, the velocity $<\dot{\phi}>$ does not
redshift to zero as the universe expands, but it stays constant,
and indeed the energy momentum tensor is identical of that of a
cosmological constant. However, the ghost condensate is a physical
fluid, and so, it has physical fluctuations which can be defined
as:
\begin{equation}
\phi=M^2t+\pi
\end{equation}

The ghost condensate then gives an alternative way of realizing De
Sitter phases in the universe. The symmetries of the theory allow
us to construct a systematic and reliable effective Lagrangian for
$\pi$ and gravity at energies lower than the ghost cut-off $M$.
Neglecting the interactions with gravity, around flat space, the
effective Lagrangian for $\pi$ has the form:
\begin{equation}
S=\int d^4x
\frac{1}{2}\dot{\pi}^2-\frac{\alpha}{2M^2}(\nabla^2\pi)^2-\frac{\beta}{2M^2}\dot{\pi}(\nabla\pi)^2+...
\end{equation}
where $\alpha$ and $\beta$ are order one coefficients. In
\cite{Ark1}, it was shown that, in order for the ghost condensate
to be able to implement inflation, the shift symmetry of the ghost
field $\phi$ had to be broken. This could be realized adding a
potential to the ghost. The observational consequences of the
theory were no tilt in the power spectrum, a relevant amount of
non gaussianities, and the absence of gravitational waves. The non
gaussianities appeared to be the aspect closest to a possible
detection by experiments such as WMAP. Also the shape of the
3-point function of the curvature perturbation $\zeta$ was
different from the one predicted in standard inflation. In the
same paper \cite{Ark1}, the authors studied the possibility of
adding a small tilt to the ghost potential, and they did some
order of magnitude estimate of the consequences in the case the
potential decreases while $\phi$ increases.

In this paper, we perform a more precise analysis of the
observational consequences of a ghost inflation with a tilt in the
potential. We study the 2-point and 3-point functions. In
particular, we also imagine that the potential is tilted in such a
way that actually the potential increases as the value of $\phi$
increases with time. This configuration still allows inflation,
since the main contribution to the motion of the ghost comes from
the condensation of the ghost, which is only slightly affected by
the presence of a small tilt in the potential. This provides an
inflationary model in which $H$ is growing with time. We study the
2-point and 3-point function also in this case.\\

The paper is organized as follows. In section 2, we introduce the
concept of the tilt in the ghost potential; in section 3 we study
the case of negative tilt, we compute the 2-point and 3-point
functions, and we determine the region of the parameter space
which is not ruled out by observations; in section 4 we do the
same as we did in section 3 for the case of positive tilt; in
section 5 we summarize our conclusions.

\section{Density Perturbations}

In an inflationary scenario, we are interested in the quantum
fluctuations of the $\pi$ field, which, out of the horizon, become
classical fluctuations. In \cite{Ark2}, it was shown that, in the
case of ghost inflation, in longitudinal gauge, the gravitational
potential $\Phi$ decays to zero out of the horizon. So, the
Bardeen variable is simply:
\begin{equation}
\zeta=-\frac{H}{\dot{\phi}}\pi
\end{equation}
and is constant on superhorizon scales. It was also shown that the
presence of a ghost condensate modifies gravity on a time scale
$\Gamma^{-1}$, with $\Gamma\sim M^3/M^2_{Pl}$, and on a length
scale $m^{-1}$, with $m\sim M^2/M_{Pl}$. The fact that these two
scales are different is not a surprise since the ghost condensate
breaks Lorentz symmetry.

Requiring that gravity is not modified today on scales smaller
than the present Hubble horizon, we have to impose $\Gamma<H_0$,
which implies that gravity is not modified during inflation:
\begin{equation}
\Gamma\ll m\ll H
\end{equation}
This is equivalent to the decoupling limit
$M_{Pl}\rightarrow\infty$, keeping $H$ fixed, which implies that
we can study the Lagrangian for $\pi$ neglecting the metric
perturbations.

Now, let us consider the case in which we have a tilt in the
potential. Then, the zero mode equation for $\pi$ becomes:
\begin{equation}
\ddot{\pi}+3H\dot{\pi}+V'=0
\end{equation}
which leads to the solution:
\begin{equation}
\dot{\pi}=-\frac{V'}{3H}
\end{equation}
We see that this is equivalent to changing the velocity of the
ghost field.

In order for the effective field theory to be valid, we need that
the velocity of $\pi$ to be much smaller than $M^2$, so, in
agreement with \cite{Ark1}, we define the parameter:
\begin{eqnarray}
&&\delta^2=-\frac{V'}{3HM^2} \  {\rm for}\ V'<0 \\ \nonumber
&&\delta^2=+\frac{V'}{3HM^2} \  {\rm for}\ V'>0
\end{eqnarray}
to be $\delta^2\ll 1$. We perform the analysis for small tilt, and
so at first order in $\delta^2$.

At this point, it is useful to write the 0-0 component of the
stress energy tensor, for the model of \cite{Ark2}:
\begin{equation}
T_{00}=-M^4P(X)+2M^4 P'(X)\dot{\phi}^2+V(\phi)
\end{equation}
where $X=\partial_\mu\phi\partial^\mu\phi$. The authors show that
the field, with no tilted potential, is attracted to the minimum
of the function $P(X)$, such that, $P(X_{min})=M^2$. So, adding a
tilt to the potential can be seen as shifting the ghost field away
from the minimum of $P(X)$.

Now, we proceed to studying the two point function and the three
point function for both the cases of a positive tilt and a
negative tilt.

\section{Negative Tilt}

Let us analyze the case $V'<0$.\\
\subsection{2-Point Function}
To calculate the spectrum of the $\pi$ fluctuations,
we quantize the field as usual:
\begin{equation}
\pi_k(t)=w_k(t)\hat{a}_k+w^*_k(t)\hat{a}^\dag_{-k}
\end{equation}
The dispersion relation for $w_k$ is:
\begin{equation}
\omega^2_k=\alpha \frac{k^4}{M^2}+\beta \delta^2k^2
\end{equation}
Note, as in \cite{Ark2}, that the sign of $\beta$ is the same as
the sign of $<\dot{\phi}>=M^2$. In all this paper we shall
restrict to $\beta\geq 0$, and so the sign of $\beta$ is fixed.

We see that the tilt introduces a piece proportional to $k^2$ in
the dispersion relation. This is a sign that the role of the tilt
is to transform ghost inflation to the standard slow roll
inflation. In fact, $\omega^2\sim k^2$ is the usual dispersion relation
for a light field.

Defining $w_k(t)=u_k(t)/a$, and going to conformal time
$d\eta=dt/a$, we get the following equation of motion:
\begin{equation}
u_k''+(\beta\delta^2 k^2+\alpha\frac{k^4 H^2
\eta^2}{M^2}-\frac{2}{\eta^2})u_k=0 \label{diffeq}
\end{equation}
If we were able to solve this differential equation, than we could
deduce the power spectrum. But, unfortunately, we are not able to
find an exact analytical solution. Anyway, from (\ref{diffeq}), we
can identify two regimes: one in which the term $\sim k^4$
dominates at freezing out, $\omega\sim H$, and one in which it is the
term in $\sim k^2$ that dominates at that time. Physically, we
know that most of the contribution to the shape of the
wavefunction comes from the time around horizon crossing. So, in
order for the tilt to leave a signature on the wavefunction, we
need it to dominate before freezing out.There will be an
intermediate regime in which both the terms in $k^2$ and $k^4$
will be important around horizon crossing, but we decide not to
analyze that case as not too much relevant to our discussion. So,
we restrict to:
\begin{equation}
\delta^2\gg\delta^2_{cr}\equiv\frac{\alpha^{1/2}}{\beta}\frac{H}{M}
\end{equation}
where $cr$ stays for $crossing$. In that case, the term in $k^2$
dominates before freezing out, and we can approximate the
differential equation (\ref{diffeq}) to:
\begin{equation}
u_k''+(\bar{k}^2-\frac{2}{\eta^2})u_k=0
\end{equation}
where $\bar{k}=\beta^{1/2} \delta k$. Notice that this is the same
differential equation we would get for the slow roll inflation
upon replacing $k$ with $\bar{k}$.

Solving with the usual vacuum initial condition, we get:
\begin{equation}
w_k=-H\eta\frac{e^{-i\bar{k}\eta}}{2^{1/2}\bar{k}^{1/2}\eta}(1-\frac{i}{\bar{k}\eta})
\label{wafefunction}
\end{equation}
which leads to the power spectrum:
\begin{equation}
P_{\pi}=\frac{k^3}{2\pi^2}
 |w_k(\eta\rightarrow 0)|^2=\frac{H^2}{4\pi^2\beta^{3/2}\delta^3}
\end{equation}
and, using $\zeta=-\frac{H}{\dot{\phi}}\pi$,
\begin{equation}
P_{\zeta}=\frac{H^4}{4\pi^2 \beta^{3/2}\delta^3 M^4}
\end{equation}
This is the same result as in slow roll inflation, replacing $k$
with $\bar{k}$. Notice that, on the contrary with respect to
standard slow roll inflation, the denominator is not suppressed by
slow roll parameters, but by the $\delta^2$ term.

The tilt is given by:
\begin{eqnarray}
&&n_s-1\equiv\frac{d ln(P_\zeta)}{d ln(k)}=(4\frac{d ln(H)}{d ln
k}-\frac{3}{2}\frac{dln\beta}{d ln
k}-\frac{3}{2}\frac{dln\delta^2}{d ln k}-2\frac{d ln \dot{\phi}}{d
ln k})|_{k=\frac{aH}{\beta^{1/2}\delta}}=\\ \nonumber
&&=\frac{2M^2V'}{HV}+\frac{V''}{H^2}\Big(\frac{1}{2\delta^2}+\frac{2}{9}+\frac{4M^4}{V}(1-2P''M^8)\Big)
\end{eqnarray}
where $k=\frac{aH}{\beta^{1/2}\delta}$ is the momenta at freezing
out, and where $P$ and its derivatives are evaluated at $X_{min}$.

Notice the appearance of the term $\sim\frac{1}{\delta^2}$, which
can easily be the dominant piece. Please remind that, anyway, this
is valid only for $\delta^2\gg\delta^2_{cr}$. Notice also that,
for the effective field theory to be valid, we need:
\begin{equation}
\frac{V'}{3H}\ll M^2
\end{equation}
so, $\frac{M^2V'}{HV}\ll\frac{M^4}{V}$. This last piece is in
general $\ll 1$ if the ghost condensate is present today. In order
to get an estimate of the deviation from scale invariance, we can
see that the larger contribution comes from the piece in
$\sim\frac{V''}{\delta^2 H^2}$. From the validity of the effective
field theory, we get:
\begin{equation}
\delta^2M^2H\cong|V'|\gtrsim|V''|\Delta\phi=|V''|(M^2/H)N_e\Rightarrow|V''|<\delta^2
\frac{H^2}{N_e}
\end{equation}
where $N_e$ is the number of e-foldings to the end of inflation.
So, we deduce that the deviation of the tilt can be as large as:
\begin{equation}
|n_s-1|\leq\frac{1}{N_e}
\end{equation}
This is a different prediction from the exact $n_s=1$ in usual
ghost inflation.
\subsection{3-Point Function}
Let us come to the computation of the three point function. The
leading interaction term (or the least irrelevant one), is given
by \cite{Ark1}:
\begin{equation}
L_{int}=-\beta \frac{e^{Ht}}{2M^2}(\dot{\pi}(\nabla \pi)^2)
\end{equation}
Using the formula in \cite{Maldacena:}:
\begin{equation}
<\pi_{k_1}(t)\pi_{k2}(t)\pi_{k3}(t)>=-i\int^{t}_{t_0}dt'<[\pi_{k_1}(t)\pi_{k2}(t)\pi_{k3}(t),\int
d^3x H_{int}(t')]>
\end{equation}
we get \cite{Ark1}:
\begin{eqnarray}
&&<\pi_{k_1}\pi_{k_2}\pi_{k_3}>=\frac{i\beta}{M^2}(2\pi)^3\delta^3(\sum
k_i)\\
\nonumber
&&w_1(0)w_2(0)w_3(0)((\vec{k_2}.\vec{k_3})I(1,2,3)+cyclic+c.c)
\end{eqnarray}
where $cyclic$ stays for cyclic permutations of the $k$'s, and
where
\begin{equation}
I(1,2,3)=\int^0_{-\infty}\frac{1}{H\eta}w^*_1(\eta)'w^*_2(\eta)w^*_3(\eta)
\end{equation}
and the integration is performed with the prescription that the
oscillating functions inside the horizon become exponentially
decreasing as $\eta\rightarrow-\infty$.

We can do the approximation of performing the integral with the
wave function (\ref{wafefunction}). In fact, the typical behavior
of the wavefunction will be to oscillate inside the horizon, and
to be constant outside of it. Since we are performing the
integration on a path which exponentially suppresses the
wavefunction when it oscillates, and since in the integrand there
is a time derivative which suppresses the contribution when a
wavefunction is constant, we see that the main contribution to the
three point function comes from when the wavefunctions are around
freezing out. Since, in that case, we are guaranteed that the term
in $k^2$ dominates, then we can reliably approximate the
wavefunctions in the integrand with those in (\ref{wafefunction}).
Using that $\zeta=-\frac{H}{\dot{\phi}}\pi$,we get:
\begin{eqnarray}
&&<\zeta_{k_1}\zeta_{k_2}\zeta_{k_3}>=(2\pi)^3\delta^3(\sum
k_i)\frac{H^8}{4\beta^3\delta^8M^8}\\ \nonumber
&&\frac{1}{k^3_t\prod^3_{i=1}
k^3_i}\Big(k^2_1(\vec{k_2}.\vec{k_3})\Big((k_2+k_3)k_t+k^2_t+2k_3k_2\Big)+cyclic\Big)
\label{3-point}
\end{eqnarray}
where $k_i=|\vec{k}_i|$. Let us define
\begin{equation}
F(k_1,k_2,k_3)=\frac{1}{k^3_t\prod
k^3_i}\Big(k^2_1(\vec{k_2}.\vec{k_3})\Big((k_2+k_3)k_t+k^2_t+2k_3k_2\Big)+cyclic\Big)
\end{equation}
which, apart for the $\delta$ function, holds the $k$ dependence
of the 3-point function.

The obtained result agrees with the order of magnitude estimates
given in \cite{Ark1}:
\begin{equation}
\frac{<\zeta^3>}{(<\zeta^2>)^{3/2}}\sim\frac{1}{\delta^8}(\frac{H}{M})^8\frac{1}{(\frac{1}{\delta^{3}}(\frac{H}{M})^{2})^{3/2}}\sim
\frac{1}{\delta^{7/2}}(\frac{H}{M})^2 \label{estimate}
\end{equation}
The total amount of non gaussianities is decreasing with the tilt.
This is in agreement with the fact that the tilt makes the ghost
inflation model closer to slow roll inflation, where, usually, the
total amount of non gaussianities is too low to be detectable.

The 3-point function we obtained can be better understood if we do
the following observation. This function is made up of the sum of
three terms, each one obtained on cyclic permutations of the
$k$'s. Each of these terms can be split into a part which is
typical of the interaction and of scale invariance, and the rest
which is due to the wave function. For the first cyclic term, we
have:
\begin{equation}
Interaction=\frac{(\vec{k_2}.\vec{k_3})}{k_1 k^3_2k^3_3}
\end{equation}
while, the rest, which I will call wave function, is:
\begin{equation}
Wavefunction=\frac{((k_2+k_3)k_t+k^2_t+2k_2k_3)}{k^3_t}
\end{equation}
The interaction part appears unmodified also in the untilted ghost
inflation case. While the wave function part is characteristic of
the wavefunction, and changes in the two cases.

Our 3-point function can be approximately considered as a function
of only two independent variables. The delta function, in fact,
eliminates one of the three momenta, imposing the vectorial sum of
the three momenta to form a closed triangle. Because of the
symmetry of the De Sitter universe, the 3-point function is scale
invariant, and so we can choose $|\vec{k}_1|=1$. Using rotation
invariance, we can choose $\vec{k}_1=\hat{e}_1$, and impose
$\vec{k}_2$ to lie in the $\hat{e}_1,\hat{e}_2$ plane. So, we have
finally reduced the 3-point function from being a function of 3
vectors, to be a function of 2 variables. From this, we can choose
to plot the 3-point function in terms of
$x_i\equiv\frac{k_i}{k_1}$, $i=1,2$. The result is shown in
fig.\ref{fig1}. Note that we chose to plot the three point
function with a measure equal to $x_2^2x_3^2$. The reason for this
is that this results in being the natural measure in the case we
wish to represent the ratio between the signal associated to the
3-point function with respect to the signal associated to the
2-point function \cite{cremzalda}. Because of the triangular
inequality, which implies $x_3\leq 1-x_2$, and in order to avoid
to double represent the same momenta configuration, we set to zero
the three point function outside the triangular region: $1-x_2\leq
x_3\leq x_2$. In order to stress the difference with the case of
standard ghost inflation, we plot in fig.\ref{fig2} the
correspondent 3-point function for the case of ghost inflation
without tilt. Note that, even though the two shapes are quite
similar, the 3-point function of ghost inflation without tilt
changes signs as a function of the $k$'s, while the 3-point
function in the tilted case has constant sign.
\begin{figure}[!h]
\begin{center}
\includegraphics{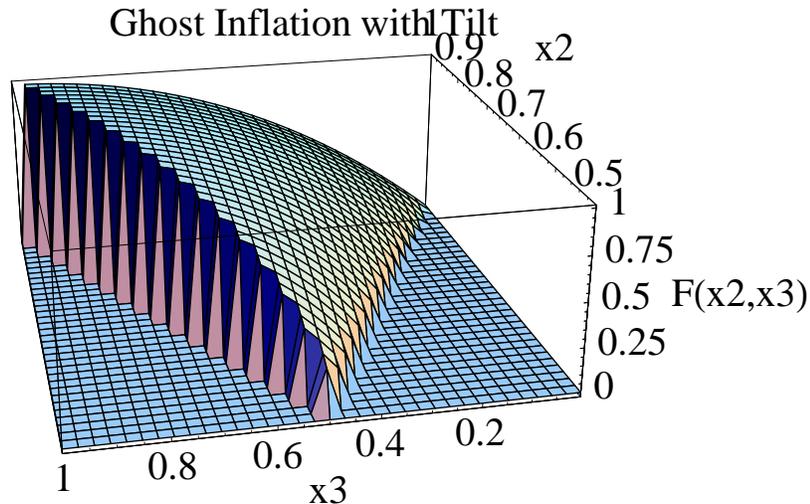}
\caption{Plot of the function $F(1,x_2,x_3)x^2_2x^2_3$ for the
tilted ghost inflation 3-point function. The function has been
normalized to have value 1 for the equilateral configuration
$x_2=x_3=1$, and it has been set to zero outside of the region
$1-x_2\leq x_3\leq x_2$}\label{fig1}
\end{center}
\end{figure}

\begin{figure}[!h]
\begin{center}
\includegraphics{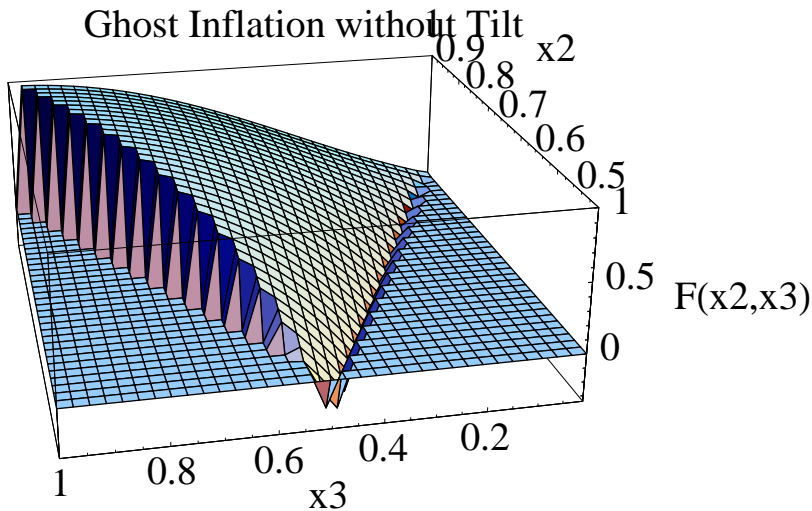}
\caption{Plot of the similarly defined function
$F(1,x_2,x_3)x^2_2x^2_3$ for the standard ghost inflation 3-point
function. The function has been normalized to have value 1 for the
equilateral configuration $x_2=x_3=1$, and it has been set to zero
outside of the region $1-x_2\leq x_3\leq x_2$
\cite{cremzalda}}\label{fig2}
\end{center}
\end{figure}

An important observation is that, in the limit as $x_3\rightarrow
0$ and $x_2\rightarrow 1$, which corresponds to the limit of very
long and thin triangles, we have that the 3 point function goes to
zero as $\sim\frac{1}{x_3}$. This is expected, and in contrast
with the usual slow roll inflation result $\sim\frac{1}{x^3_3}$.
The reason for this is the same as the one which creates the same
kind of behavior in the ghost inflation without tilt \cite{Ark1}.
The limit of $x_3\rightarrow 0$ corresponds to the physical
situation in which the mode $k_3$ exits from the horizon, freezes
out much before the other two, and acts as a sort of background.
In this limit, let us imagine a spatial derivative acting on
$\pi_{3}$, which is the background in the interaction Lagrangian.
The 2-point function $<\pi_{1}\pi_{2}>$ depends on the position on
the background wave, and, at linear order, will be proportional to
$\partial_i\pi_{3}$. The variation of the 2-point function along
the $\pi_{3}$ wave is averaged to zero in calculating the 3-point
function $<\pi_{k_1}\pi_{k_2}\pi_{k_3}>$, because the spatial
average $<\pi_{3}\partial_i\pi_{3}>$ vanishes. So, we are forced
to go to the second order, and we therefore expect to receive a
factor of $k^2_3$, which accounts for the difference with the
standard slow roll inflation case. In the model of ghost
inflation, the interaction is given by derivative terms, which
favors the correlation of modes freezing roughly at the same time,
while the correlation is suppressed for modes of very different
wavelength. The same situation occurs in standard slow roll
inflation when we study non-gaussianities generated by higher
derivative terms \cite{Creminelli}.

The result is in fact very similar to the one found in
\cite{Creminelli}. In that case, in fact, the interaction term
could be represented as:

\begin{equation}
L_{int}\sim
\dot{\varphi}^2(-\dot{\varphi}^2+e^{-2Ht}(\partial_i\varphi)^2)
\label{creminter}
\end{equation}
where one of the time derivative fields is contracted with the
classical solution. This interaction gives rise to a 3-point
function, which can be recast as:
\begin{eqnarray}
&&<\zeta_{k_1}\zeta_{k_2}\zeta_{k_3}>\sim\Big(\frac{(k_1^2(\vec{k_2}.\vec{k_3}))
}{\prod_i
(k_i^3) k_t^3}((k_2+k_3)k_t+k^2_t+2k_2k_3)+cyclic\Big)+\\
\nonumber &&\frac{12}{\prod_i (k_i^3) k_t^3}(k^3_1+k^3_2+k^3_3)
\end{eqnarray}
We can easily see that the first part has the same $k$ dependence
as our tilted ghost inflation. That part is in fact due to the
interaction with spatial derivative acting, and it is equal to our
interaction. The integrand in the formula for the 3-point function
is also evaluated with the same wave functions, so, it gives
necessarily the same result as in our case. The other term is due
instead to the term with three time derivatives acting. This term
is not present in our model because of the spontaneous breaking of
Lorentz symmetry, which makes that term more irrelevant that the
one with spatial derivatives, as it is explained in \cite{Ark1}.
This similarity could have been expected, because, adding a tilt
to the ghost potential, we are converging towards standard slow
roll inflation. Besides, since we have a shift symmetry for the
ghost field, the interaction term which will generate the non
gaussianities will be a higher derivative term, as in
\cite{Creminelli}.

We can give a more quantitative estimate of the similarity in the
shape between our three point function and the three point
functions which appear in other models. Following
\cite{cremzalda}, we can define the cosine between two three point
functions $F_1(k_1,k_2,k_3)$, $F_2(k_1,k_2,k_3)$, as:
\begin{equation}
cos(F_1,F_2)=\frac{F_1\cdot F_2}{(F_1\cdot F_1)^{1/2}(F_2 \cdot
F_2)^{1/2}}
\end{equation}
where the scalar product is defined as:
\begin{equation}
F_1(k_1,k_2,k_3)\cdot
F_2(k_1,k_2,k_3)=\int^1_{1/2}dx_2\int^{x_2}_{1-x_2}dx_3 x^4_2
x^4_3 F_1(1,x_2,x_3)F_2(1,x_2,x_3)
\end{equation}
where, as before, $x_i=\frac{k_i}{k_1}$. The result is that the
cosine between ghost inflation with tilt and ghost inflation
without tilt is approximately 0.96, while the cosine with the
distribution from slow roll inflation with higher derivatives is
practically one. This means that a distinction between ghost
inflation with tilt and slow roll inflation with higher derivative
terms , just from the analysis of the shape of the 3-point
function, sounds very difficult. This is not the case for
distinguishing from these two models and ghost inflation without
tilt.

Finally, we would like to make contact with the work in
\cite{eva}, on the Dirac-Born-Infeld (DBI) inflation. The leading
interaction term in DBI inflation is, in fact, of the same kind as
the one in (\ref{creminter}), with the only difference being the
fact that the relative normalization between the term with time
derivatives acting and the one with space derivatives acting is
weighted by a factor $\gamma^2=(1-v^2_p)^{-1}$, where $v_p$ is the
gravity-side proper velocity of the brane whose position is the
Inflaton. This relative different normalization between the two
terms is in reality only apparent, since it is cancelled by the
fact that the dispersion relation is $w\sim\frac{k}{\gamma}$. This
implies the the relative magnitude of the term with space
derivatives acting, and the one of time derivatives acting are the
same, making the shape of the 3-point function in DBI inflation
exactly equal to the one in slow roll inflation with higher
derivative couplings, as found in \cite{Creminelli}.

\subsection{Observational Constraints}
We are finally able to find the observational constraints that the
negative tilt in the ghost inflation potential implies.

In order to match with COBE:
\begin{equation}
P_\zeta=\frac{1}{4\pi^2\beta^{3/2}\delta^3}(\frac{H}{M})^4\cong
(4.8\ 10^{-5})^2\Rightarrow \frac{H}{M}\cong 0.018
\beta^{3/8}\delta^{3/4}
\end{equation}
From this, we can get a condition for the visibility of the tilt.
Remembering that
$\delta^2_{cr}=\frac{\alpha^{1/2}}{\beta}(\frac{H}{M})$, we get
that, in order for $\delta$ to be visible:
\begin{equation}
\delta^2\gg\delta^2_{cr}=0.018
\frac{\alpha^{1/2}\delta^{3/4}}{\beta^{5/8}}\Rightarrow
\delta^2\gg\delta^2_{visibility}=0.0016
\frac{\alpha^{4/5}}{\beta}
\end{equation}

In the analysis of the data (see for example \cite{wmap}), it is
usually assumed that the non-gaussianities come from a field
redefinition:
\begin{equation}
\zeta=\zeta_g-\frac{3}{5}f_{NL}(\zeta^2_g-<\zeta^2_g>)
\label{standard}
\end{equation}
where $\zeta_g$ is gaussian. This pattern of non-gaussianity,
which is local in real space, is characteristic of models in which
the non-linearities develop outside the horizon. This happens for
all models in which the fluctuations of an additional light field,
different from the inflaton, contribute to the curvature
perturbations we observe. In this case the non linearities come
from the evolution of this field into density perturbations. Both
these sources of non-linearity give non-gaussianity of the form
(\ref{standard}) because they occur outside the horizon. In the
data analysis, (\ref{standard}) is taken as an ansatz, and limits
are therefore imposed on the scalar variable $f_{NL}$. The angular
dependence of the 3-point function in momentum space implied by
(\ref{standard}) is given by:
\begin{equation}
<\zeta_{k_1}\zeta_{k_2}\zeta_{k_3}>=(2\pi)^3\delta^3(\sum_i
\vec{k}_i)(2\pi)^4(-\frac{3}{5} f_{NL} P^2_R)\frac{4\sum_i
k^3_i}{\prod_i 2k^3_i}
\end{equation}

In our case, the angular distribution is much more complicated
than in the previous expression, so, the comparison is not
straightforward. In fact, the cosine between the two distributions
is -0.1. We can nevertheless compare the two distributions
(\ref{3-point}) and (\ref{standard}) for an equilateral
configuration, and define in this way an "effective" $f_{NL}$ for
$k_1=k_2=k_3$. Using COBE normalization, we get:
\begin{equation}
f_{NL}=-\frac{0.29}{\delta^2}
\end{equation}
The present limit on non-gaussianity parameter from the WMAP
collaboration \cite{wmap} gives:
\begin{equation}
-58<f_{NL}<138 \ \ {\rm at\ 95\% \ C.L.} \label{wmapng}
\end{equation}
and it implies:
\begin{equation}
\delta^2>0.005
\end{equation}
which is larger than $\delta^2_{visibility}$ (which nevertheless
depends on the coupling constants $\alpha$,$\beta$).

Since for $\delta^2\gg\delta^2_{visibility}$ we do see the effect
of the tilt, we conclude that there is a minimum constraint on the
tilt: $\delta^2>0.005$.

In reality, since the shape of our 3-point function is very
different from the one which is represented by $f_{NL}$, it is
possible that an analysis done specifically for this shape of
non-gaussianities may lead to an enlargement of the experimental
boundaries. As it is shown in \cite{cremzalda}, an enlargement of
a factor 5-6 can be expected. This would lead to a boundary on
$\delta^2$ of the order $\delta^2\gtrsim0.001$, which is still in
the region of interest for the tilt.

Most important, we can see that future improved measurements of
Non Gaussianity in CMB will immediately constraint or verify an
important fraction of the parameter space of this model.\\

Finally, we remind that the tilt can be quite different from the
scale invariant result of standard ghost inflation:
\begin{equation}
|n_s-1|\lesssim\frac{1}{N_e}
\end{equation}
\section{Positive Tilt}

In this section, we study the possibility that the tilt in the
potential of the ghost is positive, $V'>0$. This is quite an
unusual condition, if we think to the case of the slow roll
inflation. In this case, in fact, the value of $H$ is actually
increasing with time. This possibility is allowed by the fact
that, on the contrary with respect to what occurs in the slow roll
inflation, the motion of the field is not due to an usual
potential term, but is due to a spontaneous symmetry breaking of
time diffeomorphism, which gives a VEV to the velocity of the
field. So, if the tilt in the potential is small enough, we expect
to be no big deviance from the ordinary motion of the ghost field,
as we already saw in section one.

In reality, there is an important difference with respect to the
case of negative tilt: a positive tilt introduces a wrong sign
kinetic energy term for $\pi$. The dispersion relation, in fact,
becomes:
\begin{equation}
\omega^2=\alpha \frac{k^4}{M^2}-\beta \delta^2 k^2
\end{equation}
The $k^2$ term is instable. The situation is not so bad as it may
appear, and the reason is the fact that we will consider a De
Sitter universe. In fact, deep in the ultraviolet the term in
$k^4$ is going to dominate, giving a stable vacuum well inside the
horizon. As momenta are redshifted, the instable term will tend to
dominate. However, there is another scale entering the game, which
is the freeze out scale $\omega(k)\sim H$. When this occurs, the
evolution of the system is freezed out, and so the presence of the
instable term is forgotten.

So, there are two possible situations, which resemble the ones we
met for the negative tilt. The first is that the term in $k^2$
begins to dominate after freezing out. In this situation we would
not see the effect of the tilt in the wave function. The second
case is when there is a phase between the ultraviolet and the
freezing out in which the term in $k^2$ dominates. In this case,
there will be an instable phase, which will make the wave function
grow exponentially, until the freezing out time, when this growing
will be stopped. We shall explore the phase space allowed for this
scenario, which occurs for
\begin{equation}
\delta^2\gg\delta^2_{cr}=\frac{\alpha^{1/2}}{\beta}\frac{H}{M}
\end{equation}
and we restrict to it.

Before actually beginning the computation, it is worth to make an
observation. All the computation we are going to do could be in
principle be obtained from the case of positive tilt, just doing
the transformation $\delta^2\rightarrow -\delta^2$ in all the
results we obtained in the former section. Unfortunately, we can
not do this. In fact, in the former case, we imposed that the term
in $k^2$ dominates at freezing out, and then solved the wave
equation with the initial ultraviolet vacua defined by the term in
$k^2$, and not by the one in $k^4$, as, because of adiabaticity,
the field remains in the vacua well inside the horizon. On the
other hand, in our present case, the term in $k^2$ does not define
a stable vacua inside the horizon, so, the proper initial vacua is
given by the term in $k^4$ which dominates well inside the
horizon. This leads us to solve the full differential equation:
\begin{equation}
u''+(-\beta \delta^2 k^2+\alpha \frac{k^4 H^2
\eta^2}{M^2}-\frac{2}{\eta^2})u=0
\end{equation}

Since we are not able to find an analytical solution, we address
the problem with the semiclassical WKB approximation. The equation
we have is a Schrodinger like eigenvalue equation, and the
effective potential is:
\begin{equation}
\tilde{V}=\beta \delta^2 k^2-\alpha \frac{k^4 H^2
\eta^2}{M^2}+\frac{2}{\eta^2}
\end{equation}
Defining:
\begin{equation}
\eta_{0}^2=\frac{\beta\delta^2 M^2}{\alpha H^2 k^2}
\end{equation}
we have the two semiclassical regions:\\
for $\eta\ll\eta_{0}$, the potential can be approximated to:
\begin{equation}
\tilde{V}\cong-\alpha \frac{k^4 H^2 \eta^2}{M^2}
\end{equation}
while, for $\eta\gg\eta_{0}$:
\begin{equation}
\tilde{V}\cong\beta \delta^2 k^2+\frac{2}{\eta^2}
\end{equation}
The semiclassical approximation tells us that the solution, in
these regions, is given by:\\
for $\eta\ll\eta_{0}$:
\begin{equation}
u\cong\frac{A_1}{(p(\eta))^{1/2}}e^{-i\int^{\eta_{cr}}_{\eta}}p(\eta')d\eta'
\end{equation}
while, for $\eta\gg\eta_{0}$:
\begin{equation}
u\cong\frac{A_2}{(p(\eta))^{1/2}}e^{\int^{\eta}_{\eta_{cr}}}p(\eta')d\eta'
\end{equation}
where $p(\eta)=(|\tilde{V(\eta)}|)^{1/2}$.\\
The semiclassical approximation fails for $\eta\sim\eta_{0}$. In
that case, one can match the two solution using a standard linear
approximation for the potential, and gets $A_2=A_1 e^{-i\pi/4}$
\cite{landau}. It is easy to see that the semiclassical
approximation is valid when $\delta^2\gg\delta^2_{cr}$.

Let us determine our initial wave function. In the far past, we
know that the solution is the one of standard ghost inflation
\cite{Ark1}:
\begin{equation}
u=(\frac{\pi}{8})^{1/2}(-\eta)^{1/2}H^{(1)}_{3/4}(\frac{H k^2
\alpha}{2 M}\eta^2)
\end{equation}
We can put this solution, for the remote past, in the
semiclassical form, to get:
\begin{equation}
u=\frac{1}{(\frac{2H\alpha
k^2}{M}(-\eta))^{1/2}}e^{i(-\frac{5}{8}\pi+\frac{\beta\delta^2M}{2H})}e^{i\frac{Hk^2\alpha
}{2M}\eta^2}
\end{equation}
So, using our relationship between $A_1$ and $A_2$, we get, for
$\eta\gg\eta_{0}$, the following wave function for the ghost
field:
\begin{equation}
w=u/a=-\frac{1}{2^{1/2}\bar{k}^{1/2}}e^{i(-\frac{7}{8}\pi+\frac{\beta\delta^2M}{2H})}e^{\frac{\beta\delta^2M}{\alpha
H }}(H\eta e^{\bar{k}\eta}+\frac{H}{i\bar{k}}e^{-\bar{k}\eta})
\end{equation}
Notice that this is exactly the same wave function we would get if
we just rotated $\delta\rightarrow i\delta$ in the solutions we
found in the negative tilt case. But the normalization would be
very different, in particular missing the exponential factor,
which can be large. It is precisely this exponential factor that
reflects the phase of instability in the evolution of the wave
function.

From this observation, the results for the 2-point ant 3-point
functions are immediately deduced from the case of negative tilt,
paying attention to the factors coming from the different
normalization constants in the wave function. \\
So, we get:
\begin{equation}
P_\zeta=\frac{1}{4\pi^2}\frac{e^{\frac{2\beta\delta^2M }{\alpha H
}}}{\beta^{3/2}\delta^3}(\frac{H}{M})^4
\end{equation}
Notice the exponential dependence on $\alpha,\beta,H/M$, and
$\delta^2$.

The tilt gets modified, but the dominating term
$\sim\frac{1}{\delta^2}$ is not modified:
\begin{eqnarray}
&&n_s-1=V'(\frac{2M^2}{HV}+\frac{2\pi\beta}{\alpha}\frac{\delta^2M}{H^2M^2_{Pl}})+\\
\nonumber
&&+\frac{V''}{H^2}(\frac{1}{2\delta^2}+\frac{2}{9}+\frac{4M^4}{V}(1-2P''M^8)-\frac{2\beta}{3\alpha}\frac{H}{M}
+\frac{\pi}{3}\frac{\delta^2 M^3}{H^3M^2_{Pl}}(2-4P''M^8))
\end{eqnarray}

For the three point function, we get:
\begin{eqnarray}
&&<\zeta_{k_1}\zeta_{k_2}\zeta_{k_3}>=(2\pi)^3\delta^3(\sum
k_i)\frac{H^8}{4\beta^3\delta^8M^8}\\ \nonumber
&&\frac{1}{k^3_t\prod
k^3_i}\Big(k^2_1(\vec{k_2}.\vec{k_3})\Big((k_2+k_3)k_t+k^2_t+2k_3k_2\Big)+cyclic\Big)e^{6\frac{\beta\delta^2M}{\alpha
H}}
\end{eqnarray}
which has the same $k$'s dependence as in the former case of
negative tilt. Estimating the $f_{NL}$ as in the former case, we
get:
\begin{equation}
f_{NL}\sim-\frac{0.29}{\delta^2}e^{6\frac{\beta\delta^2M}{\alpha
H}}
\end{equation}
Notice again the exponential dependence.

Combining the constraints from the 2-point and 3-point functions,
it is easy to see that a relevant fraction of the parameter space
is already ruled out. Anyway, because of the exponential
dependence on the parameters $\delta^2$,$\frac{H}{M}$, and the
coupling constants $\alpha$, and $\beta$, which allows for big
differences in the observable predictions, there are many
configurations that are still allowed.

\section{Conclusions}
We have presented a detailed analysis of the consequences of
adding a small tilt to the potential of ghost inflation.

In the case of negative tilt, we see that the model represent an
hybrid between ghost inflation and slow roll inflation. When the
tilt is big enough to leave some signature, we see that there are
some important observable differences with the original case of
ghost inflation. In particular, the tilt of the 2-point function
of $\zeta$ is no more exactly scale invariant $n_s=1$, which was a
strong prediction of ghost inflation. The 3-point function is
different in shape, and is closer to the one due to higher
derivative terms in slow roll inflation. Its total magnitude tends
to decrease as the tilt increases. It must be underlined that the
size of these effects for a relevant fraction of the parameter
space is well within experimental reach.

In the case of a positive tilt to the potential, thanks to the
freezing out mechanism, we are able to make sense of a theory with
a wrong sign kinetic term for the fluctuations around the
condensate, which would lead to an apparent instability.
Consequently, we are able to construct an interesting example of
an inflationary model in which $H$ is actually increasing with
time. Even though a part of the parameter space is already
excluded, the model is not completely ruled out, and experiments
such as WMAP and Plank will be able to further constraint the
model.

\section*{Acknowledgments}
I would like to thank  Nima Arkani-Hamed and Paolo Creminelli, who
inspired me the problem, and without whose constant help I would
have not been able to perform this calculation.

\end{document}